\documentstyle[prd,aps]{revtex}

\begin{document}
\baselineskip=16pt
\begin{center}
{\Large \bf Perturbative expansion for master equation and its applications
}\\
\ \ \\
\ \ \\
X. X. Yi$^{a,b}$,C. Li$^a$ J. C.  Su$^c$\\
{ $^a$\it Institute of Theoretical Physics, Northeast Normal University, \\
Changchun
130024, China}\\
{$^b$\it Institute of Theoretical Physics, Academia Sinica, Peking 100080,
China}{\footnote {Corresponding address}}\\
{$^c$ \it Department of Physics, Jilin University, Changchun 130023, China}\\
\end{center}
\vskip 1cm
We construct generally applicable small-loss rate expansions for
the density operator of an open system. Successive terms of those
expansions yield characteristic loss rates for dissipation processes.
Three applications are presented in order to give further insight into the
context of those expansions. The first application, of a two-level atom coupling
to a bosonic environment, shows the procedure and the advantage of the expansion,
whereas the second application that consists of a single mode
field in a cavity with linewidth $\kappa$ due to partial transmission
through one mirror illustrates a practical use of those expansions
in quantum measurements, and the third one, for an atom coupled to
modes of a lossy cavity shows the another use of the perturbative
expansion. 

The study of open quantum systems has recently attracted the attention
of physicists from various fields: cosmology[1], condensed matter[2],
quantum optics[3-7], particle physics[8], quantum measurement[9,10], and quantum
computation[11,12]. The problem can be described generally as
interest in the effective dynamics of one subsystem of several
interacting subsystems. A formal framework to describe the effective dynamics of
such a subsystem is set up in ref.[13], and a {\it short-time perturbative
expansion} for coherence loss has  also been constructed[14].
To some extent(for example, if we are interested in a behavior for finite time), 
however, time is not as good as
the loss rate as a perturbative parameter.
Motivated by this and recent experimental developments[15-20] as well as
the analysis of models related to them[21-24], we
construct generally {\it small-loss rate expansions} for
dissipation. The results suggest that these
 are useful in many areas such as high-Q Cavity QED[15-20], quantum
computation[11,12,21-26], quantum measurement[27,28], quantum optics[3-7] etc.

We consider an open quantum system, the total Hamiltonian describing such a 
system is expressed as
\begin{equation}
H=H_0+H_{env}+H_I,
\end{equation}
where $H_0$ and $H_{env}$ indicate the free Hamiltonian of the
system and of the environment, respectively. $H_I$ is the interaction
Hamiltonian between the system and the environment. It is well known that the
form of the master equation depends on the precise kind of the system-environment
interaction. In order to derive a master equation for a quite general
$H_I$, let us suppose that, in the Schr\"odinger picture, $H_I$ can
be written as [3,4]
\begin{equation}
H_I=\hbar\sum_m(X_m^+A_m+X_m^-A_m^{\dag})
\end{equation}
where the $X^{\pm}_m $are eigenoperators of the system satisfying
\begin{equation}
[H_0,X^{\pm}_m]=\pm\hbar\omega_m X_m^{\pm}.
\end{equation}
This form is quite general, since any system
operator can be decomposed into eigenoperators of $H_0$.
As shown in ref.[3,4], we can
write the  master equation in the following form
(in the Schr\"odinger picture)
\begin{eqnarray}
\dot{\rho}(t)&=&-i[H_0,\rho]+\frac 1 2 \sum_m K_m(2X_m^-\rho X^+_m-
X_m^+X_m^-\rho-\rho X_m^+X_m^-)\nonumber\\
&+&\frac 1 2 \sum_mG_m(2X_m^+\rho X^-_m-
X_m^-X_m^+\rho-\rho X_m^-X_m^+),
\end{eqnarray}
where
$$K_m=2 Re[\int_0^{\infty}d \tau e^{i\omega_m \tau}{\rm Tr}_{env}\{A_m(\tau)
A^{\dag}_m(0)\rho_{env}\}],$$
$$G_m=2 Re[\int_0^{\infty}d \tau e^{i\omega_m \tau}{\rm Tr}_{env}\{A_m^{\dag}(\tau)
A_m(0)\rho_{env}\}],$$
$\rho(t)=\rho(t,K_m,G_m)$ stands for the density operator of the system
and $\rho_{env}$ denotes the density operator of the environment.
Notice from eq.(4) that $G_m$ should vanish at zero temperature $T=0$,
while $K_m$ should not
if $A_m$ are indeed destruction operators of some kind. In
case   the constant
$G_m$ and $K_m$ are smaller than any one of the
internal coupling parameters of the
system, the density operator may be expanded in powers of $K_m$ and $G_m$[29],
\begin{eqnarray}
\rho(t,K_m,G_m)&=&\rho(t,0,0)+\sum_m\frac{\partial \rho}{\partial K_m}K_m
+\sum_m\frac{\partial \rho}{\partial G_m}G_m\nonumber\\
&+&\frac 1 2 \sum_{m,n}\frac{\partial^2\rho}{\partial K_m\partial K_n}
K_mK_n+\frac 1 2 \sum_{m,n}\frac{\partial^2\rho}{\partial G_m\partial G_n}
G_mG_n \nonumber\\
&+&\sum_{m,n}\frac{\partial^2\rho}{\partial G_m\partial K_n}
G_mK_n+...\ \ .
\end{eqnarray}
Substituting this expression into the master equation, we find the following
set of equations
\begin{eqnarray}
\dot{\rho}(t,0,0)&=&-i[H_0,\rho(t,0,0)]\\
\frac{\partial \dot{\rho}}{\partial G_m}&=&-i[H_0,\frac{\partial \rho}
{\partial G_m}]+\frac 1 2 (2X_m^+\rho(t,0,0)X_m^--X_m^-X_m^+\rho(t,0,0)
-\rho(t,0,0)X_m^-X_m^+)\nonumber\\
\frac{\partial \dot{\rho}}{\partial K_m}&=&-i[H_0,\frac{\partial \rho}
{\partial K_m}]+\frac 1 2 (2X_m^-\rho(t,0,0)X_m^+-X_m^+X_m^-\rho(t,0,0)
-\rho(t,0,0)X_m^+X_m^-)\\
\frac{\partial^2\dot{\rho}}{\partial K_m\partial K_n}&=&
-i[H_0,\frac{\partial\rho}{\partial K_m\partial K_n}]+
\frac 1 2 (2X_m^-\frac{\partial \rho }{\partial K_n}X_m^+-X_m^+X_m^-
\frac{\partial\rho}{\partial K_n}
-\frac{\partial\rho}{\partial K_n}X_m^+X_m^-)\nonumber\\
&+&
\frac 1 2 (2X_n^-\frac{\partial \rho }{\partial K_m}X_n^+-X_n^+X_n^-
\frac{\partial\rho}{\partial K_m}
-\frac{\partial\rho}{\partial K_m}X_n^+X_n^-)\nonumber\\
\frac{\partial^2\dot{\rho}}{\partial G_m\partial G_n}&=&
-i[H_0,\frac{\partial\rho}{\partial G_m\partial G_n}]+
\frac 1 2 (2X_m^+\frac{\partial \rho }{\partial G_n}X_m^--X_m^-X_m^+
\frac{\partial\rho}{\partial G_n}
-\frac{\partial\rho}{\partial G_n}X_m^-X_m^+)\nonumber\\
&+&
\frac 1 2 (2X_n^+\frac{\partial \rho }{\partial G_m}X_n^--X_n^-X_n^+
\frac{\partial\rho}{\partial G_m}
-\frac{\partial\rho}{\partial G_m}X_n^-X_n^+) \nonumber\\
\frac{\partial^2\dot{\rho}}{\partial K_m\partial G_n}&=&
-i[H_0,\frac{\partial\rho}{\partial K_m\partial G_n}]+
\frac 1 2 (2X_m^-\frac{\partial \rho }{\partial G_n}X_m^+-X_m^+X_m^-
\frac{\partial\rho}{\partial G_n}
-\frac{\partial\rho}{\partial G_n}X_m^+X_m^-)\nonumber\\
&+&
\frac 1 2 (2X_n^+\frac{\partial \rho }{\partial K_m}X_n^--X_n^-X_n^+
\frac{\partial\rho}{\partial K_m}
-\frac{\partial\rho}{\partial K_m}X_n^-X_n^+)
\end{eqnarray}
Generally speaking, given an initial condition, $\rho(0,0,0)$, we can solve  eq.(6)
exactly, which gives the zeroth order solution for the density operator $
\rho(t,0,0)$.
Substituting the zeroth order solution into eq.(7), $\partial \rho /
\partial K_m$ or $\partial \rho/ \partial G_m$ can be calculated. Following
this procedure, successive terms of the expansion (5) could be worked
out, though
the calculation is complicated.
Some words of caution are now in order. From the mathematical point of view,
the expansion (5)  holds if and only if the series
 converges. This may be satisfied easily
in physics for a large number of open systems. For example in a high-Q
cavity, the loss rate of the atom-cavity system is small enough to permit us
to expand the density operator in powers of the loss rate. At the end of
this section, we will present some discussion about this point in detail.

To illustrate the advantage of the expansion (5), we present here
a simplest model, which describes a two-level atom coupling to a bose-mode
environment. The master equation of such a system is given by[3]
\begin{equation}
\dot{\rho}=-\frac 1 2 i\Omega[\sigma_z,\rho]+\frac 1 2 \gamma\{
2\sigma^-\rho\sigma^+-\rho\sigma^+\sigma^--\sigma^+\sigma^-\rho\}
\end{equation}
with $$\gamma=2\pi Re [\int_0^{\infty}e^{i\omega_m\tau} {\rm Tr}_{env}\{b_m(\tau)
b^{\dag}_m(0)\rho_{env}\}],$$
where $b_m^{\dag}(b_m)$ stands for the creation(annihilation) operator of the
m-th mode of the environment, $\Omega$ is the Rabi frequency , and $\sigma_z(\sigma^+,
\sigma^-)$ denote the Pauli matrices. To obtain the form of the master equation
given in eq.(9), the environment was assumed to be in its vacuum state.
According to eq.(5), $\langle\sigma_z(t)\rangle$ reads:
\begin{equation}
\langle\sigma_z(t)\rangle={\rm Tr}(\rho(t,0,0)\sigma_z)+\gamma{\rm Tr}(\frac{\partial\rho}{\partial\gamma}
\sigma_z )+\frac 1 2 \gamma^2 {\rm Tr}(\frac{\partial^2\rho}
{\partial\gamma^2}\sigma_z)+...
\end{equation}
The first term in eq.(10) is $\langle\sigma_z(0)\rangle$. In order to calculate
${\rm Tr}(\frac{\partial \rho}{\partial \gamma}\sigma_z)$, we first evaluate
${\rm Tr}(\frac{\partial \dot{\rho}}{\partial \gamma}\sigma_z)$, it is given that
\begin{equation}
{\rm Tr}(\frac{\partial \dot{\rho}}{\partial \gamma}\sigma_z)
=-\langle\sigma_z(0)\rangle.
\end{equation}
Using the same procedure as mentioned, we arrive at
\begin{equation}
{\rm Tr}(\frac{\partial^2\dot{\rho}}{\partial\gamma^2}\sigma_z)=
2\langle\sigma_z(0)\rangle t.
\end{equation}
The eqs.(11,12) together give
\begin{equation}
\langle\sigma_z(t)\rangle=\langle\sigma_z(0)\rangle-\gamma t\langle\sigma_z(0)\rangle+\frac{\gamma^2}{2!}t^2
\langle\sigma_z(0)\rangle+...
\end{equation}
Using the algebra of Pauli matrices, we obtain straightforwardly from the
master equation that
\begin{equation}
\langle\sigma_z(t)\rangle=\langle\sigma_z(0)\rangle e^{-\gamma t}.
\end{equation}
A comparison between eq.(14) and (13) shows that for small $\gamma$, the
expansions are a quite good approach for the two level dissipative system, and
this result is quite general.

Noticing the eq.(13) is expanded in the product of loss rate and time, we present
here the other example to show that these expansions are generally in powers of the loss 
rate, but not in the product of time and the loss rate.
Consider the master equation given in eq.(9) 
As mentioned above, we calculate $\langle\sigma_z(t)\rangle$ in order to illustrate 
the advantage  of the expansions. The results of $\langle\sigma_z(t)\rangle$
show  no difference between the short-time expansions and the small loss-rate expansions.
To show the difference between the two expansions, we calculate 
$\langle\sigma_x(t)\rangle$.
For simplicity, we only present the results up to first order of $\gamma$.
It follows from eq.(5) that
\begin{equation}
\langle\sigma_x(t)\rangle={\rm Tr}(\rho(t,0,0)\sigma_x)+\gamma{\rm Tr}(\frac{\partial\rho}{\partial\gamma}
\sigma_x )+...
\end{equation}
It is easy to show that (setting $\hbar=1,\langle\sigma_x(0)\rangle=1\rangle$)
$${\rm Tr}(\rho(t,0,0)\sigma_x)=\cos(\Omega t).$$
In order to compute ${\rm Tr}(\frac{\partial\rho}{\partial\gamma}
\sigma_x )$, we have to calculate ${\rm Tr}(\frac{\partial\dot{\rho}}{\partial\gamma}
\sigma_x )$.
Based on the expansions, we arrive at
$${\rm Tr}(\frac{\partial\dot{\rho}}{\partial\gamma}
\sigma_x )={\rm Tr}(-i[H_0,\frac{\partial\rho}{\partial\gamma}\sigma_x])
+\frac 1 2 (2\sigma^-\rho(t,0,0)\sigma^+\sigma_x-\rho(t,0,0)\sigma^+\sigma^-\sigma_x
-\sigma^+\sigma^-\rho(t,0,0)\sigma_x)\equiv a+b.$$
Simple calculation gives
$$a=-\Omega {\rm Tr}(\frac{\partial\rho}{\partial \gamma}\sigma_y).$$
As state above, in order to compute $a$, we have to calculate
${\rm Tr}(\frac{\partial\dot{\rho}}{\partial \gamma}\sigma_y).$
Using the same procedure as above, we show that
$$a=-\Omega^2\int_0^t dt{\rm Tr}(\frac{\partial\rho}{\partial\gamma}\sigma_x)-(\cos(\Omega t)-1),$$
and
$$b=-\cos(\Omega t).$$
The results for $a$ and $b$ together give
$$\frac{\partial^2 y(t)}{\partial t^2}+\Omega^2 y(t)=2\Omega\sin(\Omega t),$$
with
$y(t)\equiv{\rm Tr}(\frac{\partial\rho}{\partial\gamma}\sigma_x)$ and 
initial conditions
$y(t=0)=0, \dot{y}(t)|_{t=0}=-1$.
This is a two order differential equation and can be solved easily, once $y(t)$ is known, 
$\langle\sigma_x(t)\rangle$ up to first order of $\gamma$ is given.
It is obvious that results given above are indeed different from the short time expansions, since
the results given by short time expansions are in powers of time $t$.\\

The results up to first order of $\gamma$(15) and an exact 
numerical results are illustrated in Fig.1. The parameters chosen are
$\Omega=2$, and time is in units of $\Omega$. In Fig.1 the scattering line represents the 
exact numerical results, whereas the dot line and the solid line show the results from the expansion. 
The dot line and the solid line are for different $\gamma$, and $\gamma$ for dot line is smaller 
than one in solid line, those curves show that the expansions (15)  are indeed a good approximation 
to the exact solution.

We need to point out that for most open systems, average values of
meaningful quantities can't be obtained exactly in any way. Therefore
the
expansions of the density operator provide a practical approach
to the exact solution.
For example, consider a single-mode field in a lossy cavity.
The density operator for that mode obeys the following master equation in the
Schr\"odinger picture[3,4],
\begin{equation}
\dot{\rho}=-i[\omega_f a^{\dag}a,\rho]+\frac{\kappa}{2}(2a\rho a^{\dag}-
a^{\dag}a\rho-
\rho a^{\dag}a),
\end{equation}
where $\kappa$ is the linewidth of the cavity mode with
frequency $\omega_f$.
In most textbooks, the solution of the master equation is given in
terms of diagonal matrix elements $\langle n|\rho|n\rangle$ in  a
stationary state. Given
an initial condition for the density operator, the evolution of $\rho$,
however, is more useful than the stationary solution. In what follows,
we present a solution of the master equation in a number state (Fock state)
basis.

For a high-Q cavity, the linewidth $\kappa$ due to partial transmission
through one mirror is so small that we can expand $\rho$ in powers
of $\kappa$:
\begin{equation}
\rho(t)=\rho^0(t)+\frac{\partial\rho}{\partial \kappa}\kappa+\frac 1 2
\frac{\partial^2\rho}{\partial \kappa^2}\kappa^2+...,
\end{equation}
In a number state basis $\{|n\rangle, n=0,1,2,3...\}$, the
expansion can be written as
\begin{equation}
\rho(t)=\sum_{m,n}\rho^0_{mn}(t)|m\rangle\langle n|+\sum_{m,n}\sum_{k=1}^{\infty}
\frac{1}{k!}
\frac{\partial^k \rho_{mn}}
{\partial \kappa^k}|m\rangle\langle n|\kappa^k.
\end{equation}
Here the subscripts on the density operator $\rho_{mn}$ indicate matrix
elements of $\rho$ in the number basis and $\rho_0$ is the solution of eq.(16)
with $\kappa=0$. With this notation,
it follows from eqs.(6),(7) and (8) that
\begin{eqnarray}
\rho^0_{mn}(t)&=&\rho_{mn}^0(0)e^{-i(m-n)\omega_ft},\nonumber\\
\frac{\partial^k\rho_{mn}}{\partial \kappa^k}&=&c_k(t)e^{-i\omega_f(n-m)t}.
\end{eqnarray}
This iterative equation gives the density operator expansions of the system
under consideration.
Here, $\rho_{mn}(0)$ stands for the initial condition of $\rho$, and
\begin{eqnarray}
c_k(t)&=&\int_0^tF_k(t^{'})e^{i\omega_f(n-m)t^{'}}dt^{'},\nonumber\\
F_1(t)&=&\sqrt{(n+1)(m+1)}\rho^0_{m+1,n+1}(t)-\frac 1 2 m\rho^0_{mn}(t)
-\frac 1 2 n\rho^0_{mn}(t),\nonumber\\
F_k(t)&=&\sqrt{(n+1)(m+1)}\frac{\partial^{k-1}\rho_{m+1,n+1}}{\partial
\kappa^{k-1}}-\frac{m+n}{2}\frac{\partial^{k-1}\rho_{m,n}}{\partial
\kappa^{k-1}}, k=2,3,4...
\end{eqnarray}
The master equation in the form (16) is widely used in field-quadrature
measurement[27,28]. As shown in Ref.[27,28], different approximations to eq.(16)
correspond to different measurement schemes, therefore
the expansions(6-8) for the  density operator provide a new
method to develop
quantum measurement theory. In contrast to the short
time perturbative expansions[14], the expansions (6-8) hold for finite
time as long as the linewidth $\kappa$ is small. In other words, whether the 
expansions hold does not depend on time $t$.

In addition to quantum measurement, these
expansions have use in high-Q cavity QED. There are many interesting
features in cavity QED. One of them is spontaneous emission. Spontaneous
emission is so fundamental that it is usually regarded as an inherent
property of matter. The master equation for a single atom
coupling to a mode of a lossy cavity is given in the interaction
picture under
rotating-wave and dipole approximations by[15]
\begin{equation}
\dot{\rho_I}=\frac{\gamma}{2}(2\sigma^-\rho_I\sigma^+-\sigma^+\sigma^-\rho_I
-\rho_I\sigma^+\sigma^-)+\frac{\kappa}{2}(2a\rho_I a^{\dag}-a^{\dag}a\rho_I-\rho_I a^{\dag}a),
\end{equation}
where $\rho_I$ stands for the reduced density operator
of the system that consists of an atom and a cavity mode, $\gamma$ denotes the
linewidth of the atom, and $\kappa$
describes the loss rate of the cavity.This is different from eq.(9) in which
the loss of the single-mode field is neglected. It is well known that the emission
spectrum may be expressed in terms of average values of the atom operator
$\vec{\sigma}$. In this sense, we may calculate the average value of the atom
operator to replace computing the emission spectrum without any loss of
generality. Moreover, the study of many other
effects in cavity QED such as atomic dipole squeezing[30], population
trapping[31], and atomic collapse-and-revival phenomenon[32] may be reduced
to calculate and analyse the average value of the atom operator.
In the remainder of this paper, based on the expansion scheme,
we compute the average value of an atom operator given by
$A=\lambda^{(+)}\sigma^++\lambda^{(-)}\sigma^-+\lambda^{(z)}\sigma_z$.
For this end, we first of all list the expansions of the density operator
$\rho_I(t)$ in
the interaction picture
\begin{eqnarray}
\rho_I(t)&=&\rho_I^0(t)+\frac{\partial \rho_I (t)}{\partial \kappa} \kappa
+\frac{\partial \rho_I (t)}{\partial \gamma} \gamma+
\frac 1 2 \frac{\partial ^2\rho_I (t)}{\partial \kappa^2} \kappa^2
+\frac 1 2 \frac{\partial^2 \rho_I (t)}{\partial \gamma^2} \gamma^2
+\frac 1 2 \frac{\partial^2 \rho_I (t)}{\partial \gamma\partial\kappa}
\gamma\kappa
+...
\end{eqnarray}
Here,
$$\dot{\rho}_I^0(t)=0,$$
$$\frac{\partial\dot{\rho_I}(t)}{\partial \gamma}=\frac 1 2
(2\sigma^-\rho_I^0(t)\sigma^+-\sigma^+
\sigma^-\rho_I^0(t)-\rho_I^0(t)\sigma^+\sigma^-),$$
$$\frac{\partial\dot{\rho_I}(t)}{\partial \kappa}=\frac 1 2
(2a\rho_I^0(t)a^{\dag}-a^{\dag}
a\rho_I^0(t)-\rho_I^0(t)a^{\dag}a),$$
$$\frac{\partial^2\dot{\rho_I}(t)}{\partial \gamma^2}=\frac 1 2
(2\sigma^-\frac{\partial\rho_I}{\partial \gamma}\sigma^+-\sigma^+
\sigma^-\frac{\partial \rho_I}{\partial \gamma}-\frac{\partial\rho_I}
{\partial \gamma}\sigma^+\sigma^-),$$
$$\frac{\partial^2\dot{\rho_I}(t)}{\partial \kappa^2}=\frac 1 2
(2a\frac{\partial\rho_I}{\partial \kappa}a^{\dag}-a^{\dag}
a\frac{\partial \rho_I}{\partial \kappa}-\frac{\partial\rho_I}
{\partial \kappa}a^{\dag}a),$$
$$\frac{\partial^2\dot{\rho_I}(t)}{\partial \gamma\partial\kappa}=\frac 1 2
(2\sigma^-\frac{\partial\rho_I}{\partial \kappa}\sigma^+-\sigma^+
\sigma^-\frac{\partial \rho_I}{\partial \kappa}-\frac{\partial\rho_I}
{\partial \kappa}\sigma^+\sigma^-)\\
+
\frac 1 2
(2a\frac{\partial\rho_I}{\partial \gamma}a^{\dag}-a^{\dag}
a\frac{\partial \rho_I}{\partial \gamma}-\frac{\partial\rho_I}
{\partial \gamma}a^{\dag}a).
$$
It is easy to show that for any atom operator $A$,
\begin{equation}
{\rm Tr}(\frac{\partial^n\dot{\rho}_I(t)}{\partial \kappa^n}A)={\rm Tr}(\frac
{\partial^n\dot{\rho}_I(t)}{\partial \kappa ^m\partial\gamma^{n-m}}A)=0
\end{equation}
for
$n\geq m\neq 0$, while
\begin{eqnarray}
{\rm Tr}(\frac{\partial\dot{\rho}_I}{\partial\gamma}A)&=&
\frac 1 2 {\rm Tr}(\rho^0(t)B)\nonumber\\
{\rm Tr}(\frac{\partial ^n\dot{\rho}_I}{\partial
\gamma^n}A)&=&\frac 1 2 {\rm Tr}(\frac{\partial ^{n-1} \rho_I}{\partial
\gamma^{n-1}}B),
\end{eqnarray}
where $B=-4\lambda^{(+)}\sigma^+-4\lambda^{(-)}\sigma^-.$ Eqs.(23)
and (24) suggest that the average value for any atom operator can
be calculated analytically as an expansion
in powers of $\kappa$ and $\gamma$, provided $\rho^0(t)$ (the zeroth order
 density operator in the Schr\"odinger picture) is known. Generally speaking,
 given a initial condition for $\rho$, the $\rho^0(t)$ that obeys
the von Neumann equation can be given readily. In the model presented above,
 the von Neumann equation is given by
 \begin{equation}
\dot{\rho}^0=-i[H_0,\rho^0],
\end{equation}
Where $H_0$ denotes the free Hamiltonian for the cavity-atom system
(Jaynes-Cummings model)
\begin{equation}
H_0=\omega_fa^{\dag}a+\frac 1 2 \omega_a\sigma_z+g(a^{\dag}\sigma^-+\sigma^+a).
\end{equation}
If the cavity-atom system is initially in a state
$|e,n\rangle=|e\rangle\otimes|n\rangle$, i.e. the atom is in its excited state, while the
single-mode cavity is in the number state $|n\rangle$, then $\rho^0(t)$ reads
\begin{equation}
\rho^0(t)=|\psi^0(t)\rangle\langle\psi^0(t)|,
\end{equation}
where
\begin{eqnarray}
|\psi^0(t)\rangle&=&\frac 1 2 \sin\theta_{n+1}(e^{-iE_+(n+1)t}-e^{-iE_-(n+1)t})|g,n+1\rangle\nonumber\\
&+&(\sin^2\frac{\theta_{n+1}}{2}e^{-iE_+(n+1)t}+\cos^2\frac{\theta_{n+1}}{2}
e^{-iE_-(n+1)t})|e,n\rangle.
\end{eqnarray}
\begin{equation}
E_{\pm}(n+1)=\frac{\omega_f}{2}(2n+1)\pm\sqrt{\delta^2+g^2(n+1)},
\theta_{n+1}=arctg\frac{2g\sqrt{n+1}}{\delta},\delta=\omega_f-\omega_a.
\end{equation}
It follows from eq.(6,7,8) that
\begin{eqnarray}
{\rm Tr}(\rho^0(t)A)&=&\frac 1 4{\rm Tr}(\rho_0(t)B)+
\lambda^{(z)}\sum_n\{ |\sin^2\frac{\theta_{n+1}}{2}e^{-iE_+(n+1)t}+
\cos^2\frac{\theta_{n+1}}{2}e^{-iE_-(n+1)t}|^2\nonumber\\
&-&\sin^2\theta_{n+1}\sin^2(\sqrt{\delta^2+g^2(n+1)}t)\}\nonumber\\
{\rm Tr}(B\rho^0(t))&=&\sum_n\{ -2\lambda^{(+)}\sin\theta_n(
e^{-iE_+(n)t}-e^{-iE_-(n)t})\nonumber\\
&\ \ &(\sin^2\frac{\theta_{n+1}}{2}
e^{iE_+(n+1)t}+\cos^2\frac{\theta_{n+1}}{2}e^{iE_-(n+1)t})-\nonumber\\
&\ \ &2
\lambda^{(-)}(\sin^2\frac{\theta_{n+1}}{2}e^{-iE_+(n+1)t}+\cos^2
\frac{\theta_{n+1}}{2}e^{-iE_-(n+1)t})\nonumber\\
&\ \ &
\sin\theta_n(e^{iE_+(n)t}-e^{iE_-(n)t}) \}.
\end{eqnarray}
Then successive perturbative
terms of average value up to second order of $\gamma$
for an atom operator are given by
\begin{equation}
\langle A\rangle(t)={\rm Tr}(\rho_0(t)A)+\frac{\gamma}{2}\int_0^t
{\rm Tr}(\rho^0(t^{'})B)dt^{'}-\frac{\gamma^2}{2}\int_0^t dt^{'}\int_0^{t^{'}}{\rm Tr}
(\rho^0(t^{''})B)dt^{''}+....
\end{equation}
Based on the short-time expansion, $\langle A\rangle (t)={\rm Tr}(A\rho(t))
\simeq A_0+A_1 t+A_2 t^2+...$, in powers of $t$, where
$A_0={\rm Tr}(A\rho(0))$, $A_1={\rm Tr}(A\frac{\partial\rho}{\partial t}(0))$ and
$A_2=\frac 1 2 {\rm Tr}(A\frac{\partial ^2\rho}{\partial t^2}(0)).$
This is quite different from the results given by eq.(31).

Although we are currently investigating
the perturbative expansion for an open system,
we opt here for a few  qualitative comments. Mathematically, the perturbative
expansion is a good approach to the exact solution
of the master equation so long as
loss rates $\gamma$ and $\kappa$ are smaller than
all other internal coupling constants of the system.
This condition  holds for high-Q cavities from the physical point of view.
In fact, an optical cavity of $\sim$ 20$\mu m$ diameter has
$g/2\pi\sim 125MHz$ and $\kappa/2\pi\sim 100 KHz$ for
reasonable $Q\sim 10^9$. Thus the ratio $g/\kappa\sim 10^3$. Even
$g/\kappa\sim 10^4$ seems feasible for microspheres[33].
Generally, in the optical domain $g/\gamma\sim 10^2$, great enough
for the perturbative expansion in powers of $\gamma$ to hold.

In the end of this paper, we turn our attention to study the decoherence in $N$ two-level atoms.
This problem is usually related to the register in  quantum computer. A few
papers[11,12,34] have been published on this subject , but
a key additional feature of the present paper is to study the decoherence from
a new aspect.
If the system consists of $N$ two-level atoms, the decoherence is due to the
inevitable coupling of the $N$ atoms to the external environment.
Generally, the environment may be treated as 
that consists of an infinite number of  oscillators. The 
Hamiltonian describing such decoherence process takes the form
\begin{eqnarray}
H&=&H_s+H_{env}+H_I,\nonumber\\
H_s&=&\sum_{i=1}^{N}\Omega_i\sigma_i^z,\nonumber\\
H_{env}&=&\sum_{k=1}^{\infty}\omega_kb_k^{\dag}b_k,\nonumber\\
H_I&=&\sum_{i=1}^{N}\sum_{k=1}^{\infty}(g_{ki}b_k^{\dag}\sigma_i^-+h.c.),
\end{eqnarray}
where  $\sigma_i^{\alpha}$ are the spin-$\frac 1 2 $ Pauli operators ($i$
denotes the qubit index) and $b_k, b_k^{\dag}$ are the bosonic operators,
$H_s$, $H_{env}$ are the free Hamiltonian of the system and the environment,
respectively. And  $H_I$ stands for the $N$ qubits-environment interaction.
This model is closely related to the Dicke maser model[35,36]. The Hamiltonian
(32) is
complicated so it is hard to find its exact solution though the Hilbert space
associated with this model can split into invariant eigenspaces[34].
Fortunately, with the perturbative approach created in the previous section,
the complex system can be easily treated. To start with,  we
give the master equation of the system
\begin{eqnarray}
\dot{\rho}&=&-i[H,\rho]+\frac 1 2 \{\sum_i K_i (2\sigma_i^-\rho\sigma_i^+
-\rho\sigma_i^+\sigma_i^--\sigma_i^+\sigma_i^-\rho)\}\nonumber\\
&+&\frac 1 2 \{ \sum_i G_i(2\sigma_i^+\rho\sigma_i^--\sigma_i^-\sigma_i^+
\rho-\rho\sigma_i^-\sigma_i^+)\}\nonumber\\
&=&-i[H,\rho]+\cal {L}\rho.
\end{eqnarray}
Here
$$K_m=2Re\int d\tau e^{i\Omega_m\tau} {\rm Tr}_{env}\{A_m(\tau)A_m^{\dag}(0)
\rho_{env}\},$$
$$G_m=2Re\int d\tau e^{i\Omega_m\tau} {\rm Tr}_{env}\{A_m^{\dag}(\tau)A_m(0)
\rho_{env}\},$$
$$A_m(\tau)=\sum_{j=1}^{\infty}\hbar g_{mj}b_je^{-i\omega_j\tau}.$$
In the case of so-called Dicke limit[35,36], $A_m$ does not depend
on the atom index $m$. This holds, for example, when the typical environment wavelengths
are much greater than the distances between
the atoms.

In order to study the decoherence of the atoms, we assume that
the initial state of the system is
\begin{equation}
|\psi_m\rangle=S_m^+|0\rangle+\prod_{j=m+1}^N|0\rangle),
\end{equation}
where $S_m^+=\prod_{j=1}^{m\leq N}\sigma_j^+|0\rangle$, and $|0\rangle=|0\rangle_1\otimes|0\rangle_2
\otimes...\otimes|0\rangle_N$ stands for the lower state of the atoms. Eq.(34)
indicates that there are $m$ atoms in the
upper state $|1\rangle$, and the rest of
the $N$ atoms are in their lower state. With those initial conditions,
the probability of the atoms remaining in the initial state is given by $F_m(t)$
\begin{equation}
F_m(t)=1-\frac{K}{\Gamma_{K1}}-\frac{K^2}
{\Gamma_{K2}}-\frac{G}{\Gamma_{G1}}-\frac{G^2}{\Gamma_{G2}}-\frac{GK}{\Gamma_{GK}}
+O(G^3)+O(K^3),
\end{equation}
with
$$\frac{1}{\Gamma_{K1}}=t[2+2(2m-N)],$$
$$\frac{1}{\Gamma_{K2}}=-t^2[2+2(2m-N)],$$
$$\frac{1}{\Gamma_{G1}}=t[2-2(2m-N)],$$
$$\frac{1}{\Gamma_{G2}}=-t^2[2-2(2m-N)],$$
$$\frac{1}{\Gamma_{GK}}=\frac 1 2GKt^2[4(2m-N)-4].$$
Here, we suppose that all qubits are alike, so $\Omega_m=\Omega$ and
$K_m=K$ and $G_m=G$.
$G$ and $K=G+1$ depend on environment temperature $T$ through
$G=1/[exp(\hbar\Omega/kT)-1]$, which indicate that the probability
decrease with the temperature increasing.
In fact, the fidelity in the field of quantum information is nothing
but an overlap between the initial and final state of the qubits(two-level system).
Eq. (35) suggests that the fidelity depends on  $m$ i.e. the number of the atoms in  upper state
initially.
And to  get the maximum of the fidelity, the variable $m$ should be
taken as small as possible.

To sum up, in this paper, we construct the small-loss rate perturbative
expansion for the density operator of an open system. The
expansions provide a quite good approach to the exact solution in
case  the master equation of the  system can not be solved exactly. 
As an interesting application of this expansions, we used it to 
calculate some average values such as $\sigma_z$ and $\sigma_x$ in dissipative two-level
system, the expansions of the density operator for a single-mode field in
a lossy cavity are also presented, and 
the dynamical property in $N$ two-level atom system.

 In addition, the other meaningful quantities of the open system such as
 energy, occupation probability etc. can be expanded in the same spirit of
 the density operator, so long as the master equation of the system is known.

{\bf Figure captions}\\
Fig.1: $\sigma_x(t)$ vs time $t$, the parameter chosen is $\Omega=2$. 
The scattering line represent the exact numerical results, whereas the
solid line and the dot line show the results from  the expansion,
$\gamma$ in the solid line and the dot line is different,
dot line:$\gamma=0.01$,solid line:$\gamma=0.05$.
\end{document}